\documentclass[twocolumn,a4paper]{article} 
\usepackage[dvipdfmx]{graphicx}
\usepackage{authblk}
\usepackage{amsmath} 
\usepackage{txfonts}
\usepackage{color}
\usepackage{subcaption}
\usepackage{xcolor}

\newcommand{\mysubref}[2]{\ref{#1}(\subref{#1:#2})}
\newcommand{\mycite}[1]{\cite{#1}}
\newcommand{\eq}[1]{Eq.~(\ref{#1})}

\newcommand{\AL}[1]{\begin{align}#1\end{align}}

\newcommand{\LB}[1]{\label{#1}}
\newcommand{\NN}{\nonumber\\}

\newcommand{\AD}[1]{\left[#1\right]}
\newcommand{\tr}{^\top}
\newcommand{\Real}{\mathbb{R}}

\newcommand{\vect}[1]{\boldsymbol{#1}}

\newcommand{\eps}{\varepsilon}

\newcommand{\expe}[2]{\mathbb{E}_{#1}\!\AD{#2}}
\newcommand{\var}[2]{\mathrm{Var}_{#1}\!\AD{#2}}

\newcommand{\ftar}{f}
\newcommand{\targ}{p}

\newcommand{\itar}{I}
\newcommand{\titar}{\tilde{I}}

\newcommand{\xset}{\mathcal{X}}
\newcommand{\cset}{\mathcal{C}}

\newcommand{\fea}{\varphi}
\newcommand{\ph}{\fea}
\newcommand{\phm}{\fea_m}
\newcommand{\mm}{\mu_m}

\newcommand{\ats}{^{(t)}}

\newcommand{\vx}{\vect{x}}
\newcommand{\summ}{\sum_{m}}
\newcommand{\sumot}{\sum_{t=1}^{T}}

\newcommand{\mst}{M_{\mathrm{S}}}
\newcommand{\mra}{M_{\mathrm{R}}}
\newcommand{\dx}{d\vx}

\begin{document}

\title{Designing Zero-Mean Feature Functions for Multimodal Distributions}

\author[1]{Hiroshi Yamashita}
\author[1]{Hideyuki Suzuki}
\affil[1]{Graduate School of Information Science and Technology, Osaka University}

\maketitle

\abstract
To improve the accuracy of Monte Carlo estimation of expectations, a set of zero-mean feature functions, known as control variates, can be used. They can be used as feature functions for linear regression of the target function, and we can obtain an unbiased and variance-reduced estimate using its residual. One known way to construct such functions is a method using an equality called Stein's identity, but these functions are not sufficient for the case where the target distribution is multimodal. We propose a different approach to constructing these zero-mean functions based on distribution approximation and the density ratio. We demonstrate that combining the functions constructed by these two strategies can effectively reduce the estimation variance for a bimodal distribution.
\endabstract

\section{Introduction}

The Monte Carlo (MC) method \cite{Ripley87,Glasserman03} is one of the common tools in data science for handling uncertainty. It repeatedly simulates the system of interest and its uncertainty, and then aggregates the results to obtain its general properties. Specifically, in this study, we consider the problem of computing the expectation of a function over a probability distribution. The expectation to be computed is expressed as
\AL{
\itar=\int_{\Real^D} \ftar(\vx)\targ(\vx)\dx,
}
where $\targ(\vx)$ represents the target probability density function in the $D$-dimensional space $\Real^D$, and $\ftar$ represents the target function whose expectation is to be computed.
In the MC method, we draw random samples $\vx^{(1)}, \ldots, \vx^{(T)}$ from the distribution $\targ(\vx)$ using a method such as Markov Chain Monte Carlo (MCMC) \cite{Metropolis,Hastings}, and estimate $I$ by
\AL{
\titar\equiv \frac{1}{T}\sumot \ftar(\vx\ats).
}
If $\vx^{(1)}, \ldots, \vx^{(T)}$ are independent, the estimate is unbiased ($\mathbb{E}[\tilde{I}]=I$) and its variance is computed as $V[\tilde{I}] = V_{\vx\sim p}[\ftar(\vx)]/T$. Here, $\mathbb{E}[\cdot]$ and $V[\cdot]$ represent the expectation and variance, respectively, and their subscripts denote the distribution that the random variable follows.

\section{Variance Reduction of MC estimation by Control Variates}\label{sec:cv}

The accuracy of the estimation is governed by its variance. A simple way to reduce the estimation variance is to increase the number of samples $T$. However, other methods are also known as more direct ways to reduce the variance.

Let us assume that we have a set of functions $\phm(\vx)$ whose expectations $\mm=\expe{\vx\sim p}{\phm(\vx)}$ are known in advance, and whose linear combination approximates the target function as
\AL{
\ftar(\vx) = \Bigg({\summ b_m \phm(\vx)}\Bigg) + C + \eps(\vx),\label{eq:linear-model}
}
where $m$ runs over the index of the functions.
Then, using its residual $\eps(\vx)$, we can estimate $\titar$ as
\AL{
\titar&\equiv \Bigg( {\summ b_m\mm + C} \Bigg) + \frac{1}{T}\sumot \eps(\vx\ats)\NN
&=\frac{1}{T}\sumot \Big(\ftar(\vx\ats) - \summ b_m (\phm(\vx\ats) - \mm)\Big).\label{eq:cv-estimation}
}
This estimation is still unbiased, and its variance is $V[\tilde{I}] = V_{\vx\sim p}[\eps(\vx)]/T$, which is improved from the previous case when the model parameters $b_m$ are properly chosen.

The parameters can be obtained by a linear regression.
Specifically, we divide the samples into two disjoint sets $\xset_1$ and $\xset_2$, perform the linear regression only with $\xset_1$, and compute the estimate (\eq{eq:cv-estimation}) using $\xset_2$. Then, the estimation is still unbiased, and we can obtain the variance reduction.
This is known as a method of control variates (CV) \cite{Ripley87,Glasserman03,ZeroVariance,ControlFunctional}.
It is also worth noting that the zero-mean functions can also be used for the design of the herding dynamics \cite{Herding,Herding2010,NOLTA2025} to improve the sample generation.

We study how these feature functions are constructed, because the performance of this method is determined by how well the linear model approximates the target. We especially consider constructing zero-mean functions that satisfy $\expe{\vx\sim p}{\ftar(\vx)}=0$, because those with known nonzero expectations $\mm\neq 0$ are easily reduced to zero-mean functions by $\phm(\vx)\leftarrow \phm(\vx)-\mm$.

\section{Zero-mean Feature Functions Using Stein's Identity}\label{sec:stein}

It is known in the literature that $\expe{\vx\sim p}{\ph(\vx)}=0$ holds for
\AL{
\ph(\vx) = \nabla \log \targ(\vx) \cdot \vect{\psi}(\vx) + \nabla\cdot \vect{\psi}(\vx), \LB{eq:stein} 
}
where $\vect{\psi}$ is a vector-valued function under a suitable assumption.
This property is also known as Stein's identity and can be used to construct the zero-mean function $\phm$ by designing $\vect{\psi}_m$ \mycite{ControlFunctional,NOLTA2025,Stein1972ABF}.
To prove the relation, we can use the following integral over a smooth region $\Omega \subset\Real^D$:
\AL{
&\int_\Omega \targ(\vx)\ph(\vx) \dx\NN
&=\int_\Omega \targ(\vx) \nabla \log \targ(\vx) \cdot \vect{\psi}(\vx) \dx + \int_\Omega \targ(\vx) \nabla\cdot \vect{\psi}(\vx) \dx  \NN
&=\int_\Omega \nabla \cdot \big(\targ(\vx)\vect{\psi}(\vx)\big)=\int_{\partial \Omega} \targ(\vx)\vect{\psi}(\vx)\tr \vect{n}(\vx) dS,
}
where $\vect{n}(\vx)$ denotes the normal vector on the boundary $\partial \Omega$, and the last equality follows from the divergence theorem. We can derive the zero-mean property under a decay assumption on $\targ(\vx)$ as $\|\vx\|\to\infty$.

However, when the distribution $p$ has multiple separated modes and the region $\Omega$ covers only one of them, the integral can also be very small because $\targ(\vx)$ in the last expression becomes small over the boundary $\partial\Omega$. When the target function $f$ has different expectation values across modes, but all feature functions have nearly zero expectations within each mode, accurately representing this discrepancy in the linear model (\eq{eq:linear-model}) requires very large coefficients $b_m$, which leads to instability in the CV estimator \eq{eq:cv-estimation}.

\section{Zero-mean Feature Functions Based on Probability Density Ratio}\label{sec:ratio}

To avoid this instability, we need another class of zero-mean functions that have nonzero expectations for each mode at the same time.
To this end, we propose a method based on density approximation and the probability density ratio as follows.

Let us assume a reference probability distribution $R(\vx)$ and a zero-mean function $\tilde{\ph}(\vx)$ under $R$; $\expe{\vx\sim R}{\tilde{\ph}(\vx)}=0$. 
Then, we can define the function $\ph(\vx)$ with the zero-mean property $\expe{\vx\sim p}{\ph(\vx)}=0$ as
\AL{
\ph(\vx) &= w(\vx)\tilde{\ph}(\vx)\\
w(\vx)&\propto\frac{R(\vx)}{\targ(\vx)},
}
where the property follows directly from the assumptions.
Importantly, changing the constant factor of the coefficient function $w(\vx)$ does not affect the zero-mean property, so this method can be applied to the case where the target distribution $\targ(\vx)$ can only be computed up to a constant factor.

Let us further assume that the reference is a mixture of probability density functions $r_k(\vx)$, each approximating a mode of the target, as 
\AL{
R(\vx) \propto \sum_k \alpha_k r_k(\vx)
}
and construct the function $\tilde{\ph}$ as 
\AL{
\tilde{\ph}(\vx) = \sum_k c_k \frac{r_k(\vx)}{R(\vx)}, \label{eq:feature-basis}
}
where $k$ runs over the indices of the mixture components. Then, we can control the expectation of $\tilde{\ph}$ over each mode by the parameters $c_k$. At the same time, we can easily keep the overall expectation to zero, which is computed as $\expe{\vx\sim R(\vx)}{\tilde{\ph}(\vx)}=\sum_k c_k$ because of the assumption that $r_k$ are density functions.
Since we assume that $r_k$ approximates each mode, the weight $w(\vx)$ becomes nonzero in the corresponding mode's region. Then, the nonzero expectation for each mode is inherited by the obtained function $\ph$.

\section{Numerical Examples}

\begin{figure}
\centering
\includegraphics[width=0.95\columnwidth]{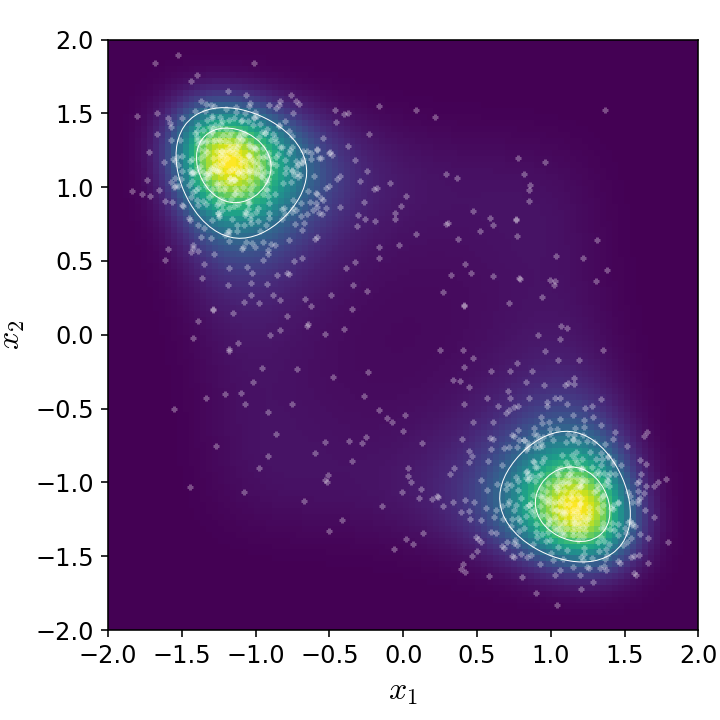}
\caption{The probability density function and the samples of the distribution (\eq{eq:example-distribution}). The density function is represented by both the color and the contour.}
\label{fig:fig1}
\end{figure}

We demonstrate how the proposed zero-mean feature functions contribute to variance reduction through numerical examples with a bimodal distribution.
Let us consider the two-dimensional distribution for $\vect{x}=(x_1, x_2)$, as illustrated in Fig.~\ref{fig:fig1}. 
This is based on an energy function $E(x_1, x_2)$, and its density function $p(x_1, x_2)$ is defined as
\AL{
p(x_1, x_2)&\propto \exp\Big( {-E(x_1, x_2)}\Big),\label{eq:example-distribution}\\
E(x_1,x_2)&=3e(x_1)+3e(x_2)+x_1x_2,\\
e(x)&=x^4 - 2x^2,
}
where $e(x)$ represents a double-well potential for a single variable, which has an energy bump at the origin. 

After generating $\xset_1$ and $\xset_2$, each consisting of 1000 samples, by standard MCMC, we randomly construct $\mst$ feature functions using Stein's identity, as described in Section~\ref{sec:stein}. 
We define the vector-valued function $\vect{\psi}_i$ as 
\AL{
\vect{\psi}_i(\vect{x}) = \sin{\Big(2\pi\big(\vect{k}_i\tr (\vect{x}/\sigma)+\theta_i\big)\Big)}\,\vect{d}_i ,
}
where $\vect{d}_i, \vect{k}_i\in\Real^2$ and $\theta_i\in[0, 2\pi)$ are randomly drawn parameters and $\sigma$ is the scale parameter. Specifically, $\vect{d}_i$ is normalized to $\|\vect{d}_i\|=1$ so that its direction is uniformly distributed, $\vect{k}_i$ is drawn from the multivariate standard normal distribution with zero mean vector and identity covariance matrix, and $\theta_i$ follows the uniform distribution on $[0, 2\pi)$. Then, we construct a zero-mean function $\ph_i$ for each $i$ by \eq{eq:stein} using $\vect{\psi}_i$ as $\vect{\psi}$. 

We also constructed $\mra$ ratio-based feature functions following the procedure described in Section~\ref{sec:ratio}. We used a reference distribution that has $\mra$ mixture components with equal weights $\alpha_k = 1$, and the $\mra$ base functions defined by \eq{eq:feature-basis}, each obtained by setting one different coefficient to $c_k = \mra - 1$ and all the others to $c_k = -1$. The mode approximations are performed by finding the parameters of the Gaussian distribution $r_k$ such that it approximates the target density for the points $\cset_k$ in the corresponding mode while keeping the ratio $r_k(\vx)/\targ(\vx)$ from exploding for the points $\cset_0$ in the remainder of the region. 

Specifically, we solved an optimization problem to minimize $\delta_k$ that satisfies 
\AL{
\big|{\log r_k(\vx) - \log \targ(\vx)}\big| &\le \delta_k
}
for all $\vx\in\cset_k$ and 
\AL{
\log r_k(\vx) - \log \targ(\vx) &\le \delta_k
}
for all $\vx\in\cset_0$. The point sets $\cset_k$ are obtained by clustering $\xset_1$ into $\mra$ clusters, and $\cset_0$ is obtained by MCMC for the broader distribution whose density is proportional to $(\targ(\vx))^{1/10}$. 
The optimization problem can be expressed and solved in the form of linear programming. 

\begin{figure}[htbp]
 \begin{subfigure}{1\columnwidth}
 \centering
 \includegraphics[width=0.95\columnwidth]{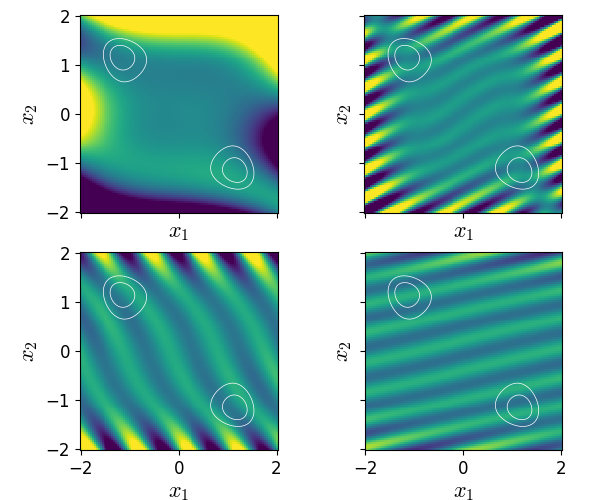}
 \caption{}\label{fig:fig2:a}
 \end{subfigure}
 \begin{subfigure}{1\columnwidth}
 \centering
\includegraphics[width=0.95\columnwidth]{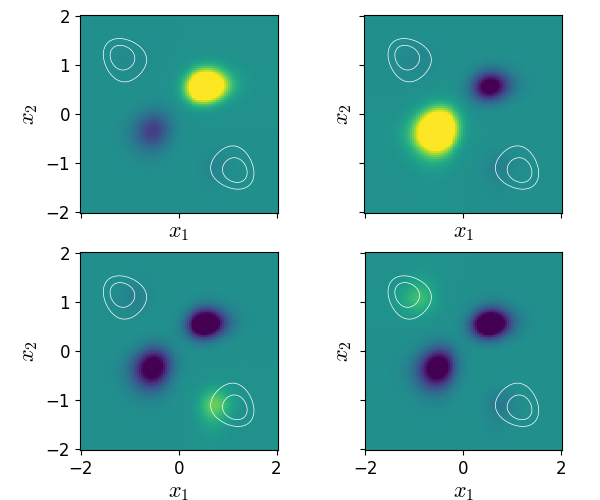}
 \caption{}\label{fig:fig2:b}
 \end{subfigure}
\caption{
    Examples of zero-mean feature functions. The contour shows the density function $\targ(\vx)$ as illustrated in Fig.~\ref{fig:fig1}. (a) Four randomly sampled feature functions constructed based on Stein's identity. The scale parameter is set as $\sigma=1$. (b) The ratio-based feature functions constructed from the distribution approximation for $\mra=4$ clusters.
}
\label{fig:fig2}
\end{figure}

\begin{figure*}[htbp]
\centering
 \begin{subfigure}{0.65\columnwidth}
 \centering
\includegraphics[width=1\columnwidth]{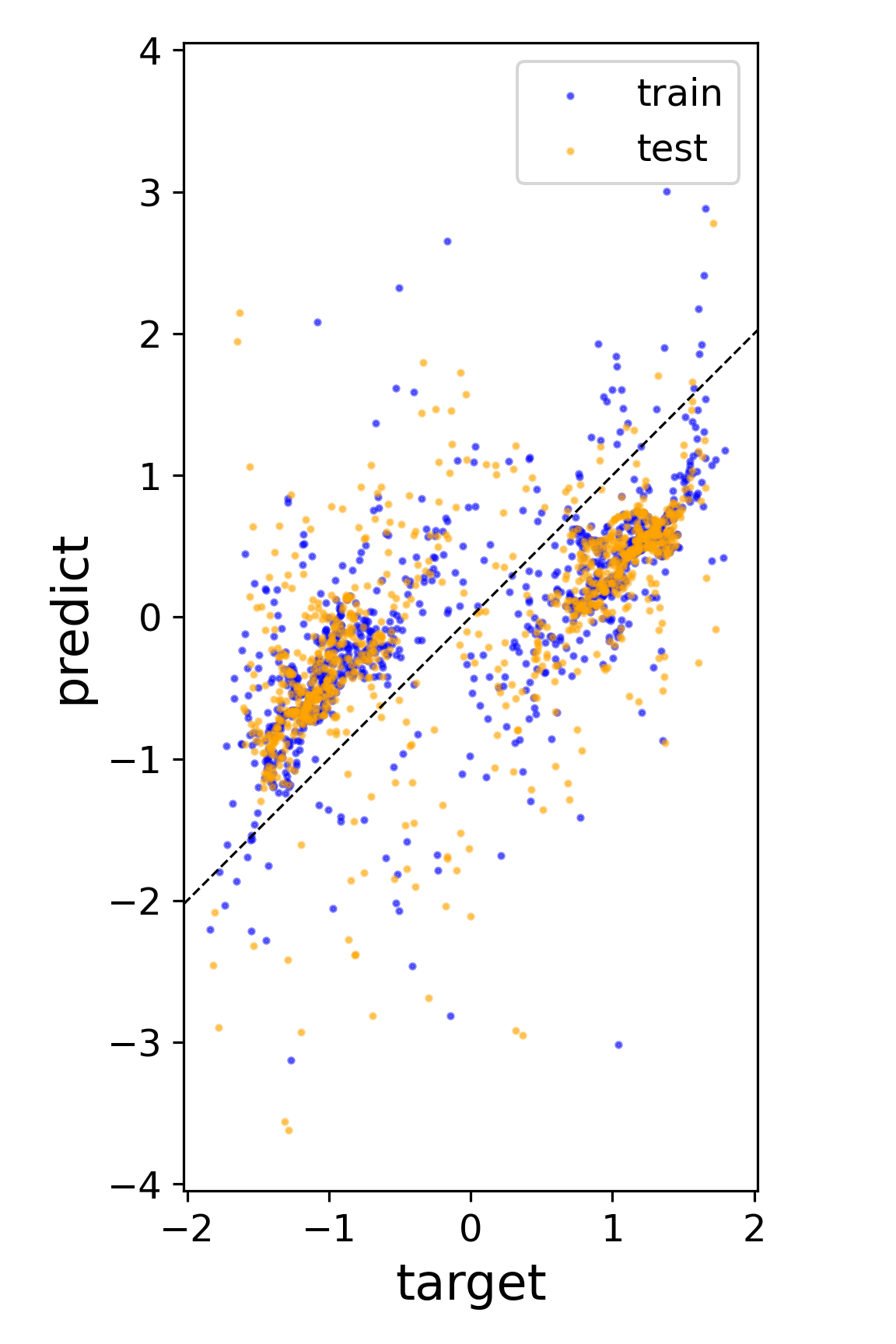}
 \caption{}
 \end{subfigure}
 \begin{subfigure}{0.65\columnwidth}
 \centering
\includegraphics[width=1\columnwidth]{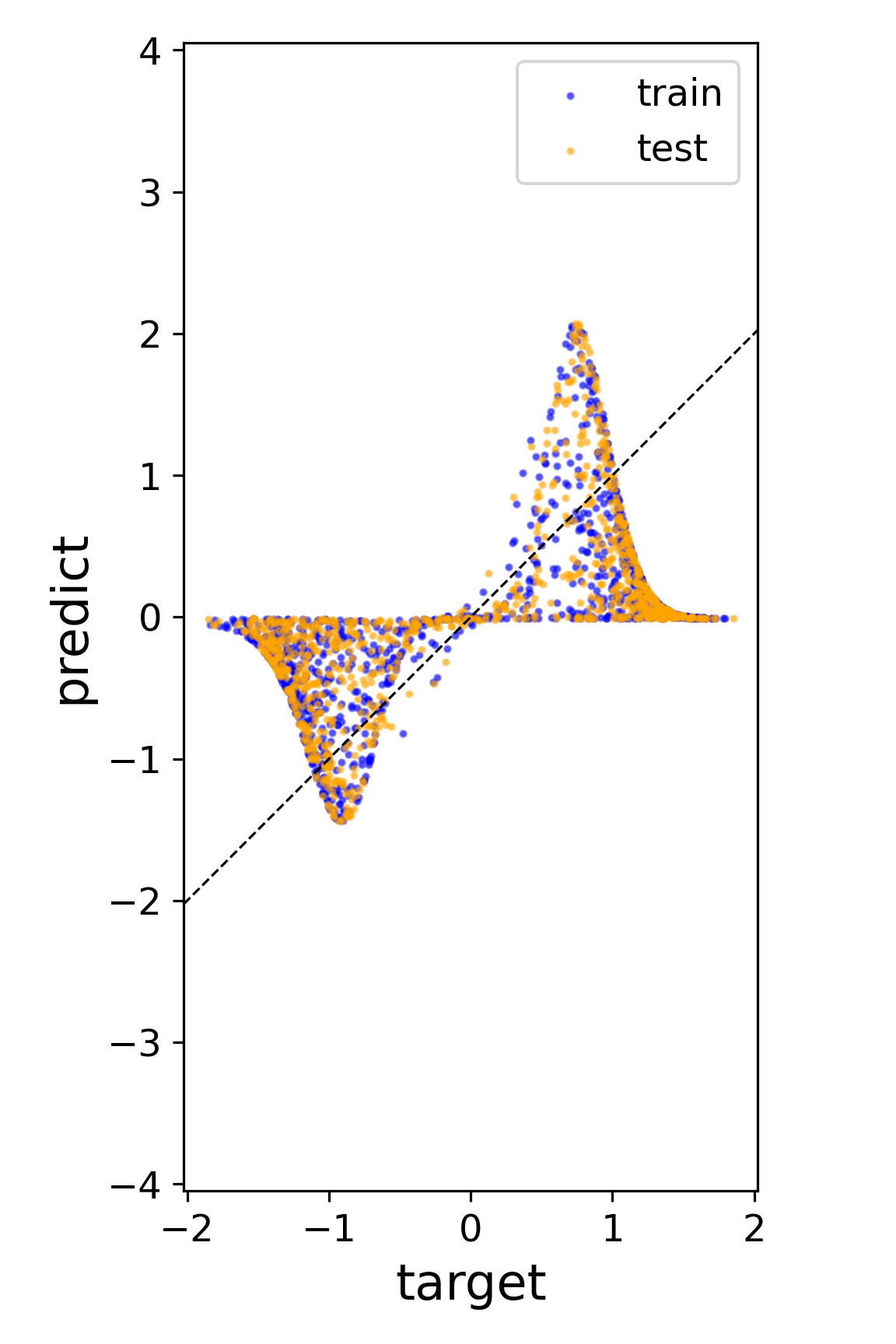}
 \caption{}
 \end{subfigure}
 \begin{subfigure}{0.65\columnwidth}
 \centering
\includegraphics[width=1\columnwidth]{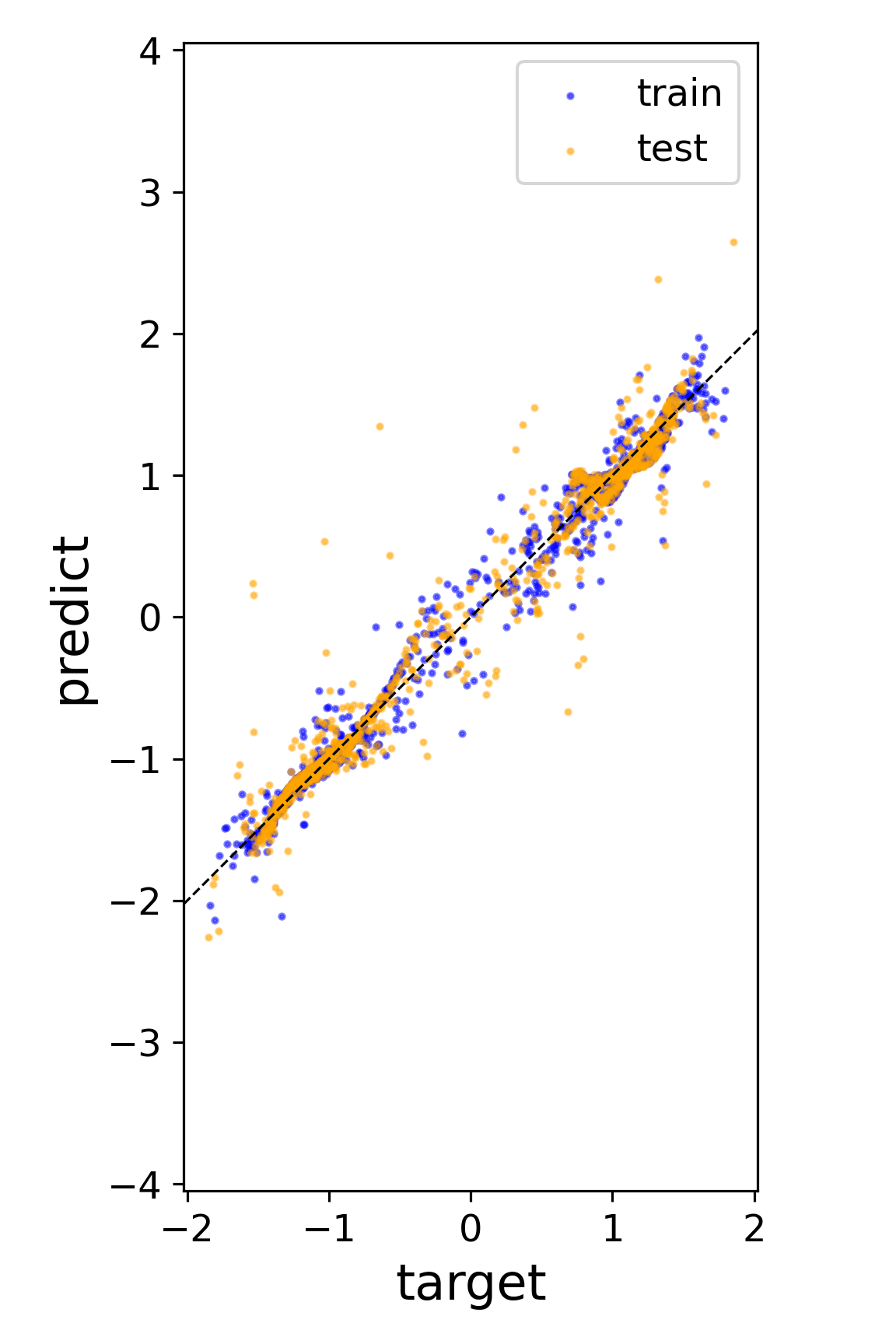}
 \caption{}
 \end{subfigure}
\caption{The predicted value in the linear model (\eq{eq:linear-model}) is plotted against its target function values $\ftar(\vx)$. The point sets where the functions are evaluated are $\xset_1$ (train) and $\xset_2$ (test). The expected accuracy of the estimation obtained by \eq{eq:cv-estimation} is represented by how the model fits the target for $\xset_2$, where the perfect fit is represented by the diagonal dotted lines in the plots. (a) Only the functions based on Stein's identity ($\sigma=1$, $\mst=100$) are used. (b) Only the proposed ratio-based functions ($\mra=4$) are used. (c) Both groups of functions are used together.}\label{fig:fig3}
\end{figure*}

Figs.~\mysubref{fig:fig2}{a} and \mysubref{fig:fig2}{b} show these two classes of zero-mean feature functions. The plots show that the functions based on Stein's identity have both positive and negative values in each mode, while the ratio-based functions have consistent signs within each mode. 
Using the target function defined as $f(x_1, x_2)=x_1$, we evaluated how the variance of the residual $\eps(\vx)$ of \eq{eq:linear-model} is reduced by the proposed method.
Fig.~\ref{fig:fig3} shows the predicted values using three combinations of the function classes. When the ratio-based functions are combined with the functions based on Stein's identity, the linear model approximates the target function well, so that the residual variance is effectively reduced.

\begin{figure}[htbp]
\centering
\includegraphics[width=0.95\columnwidth]{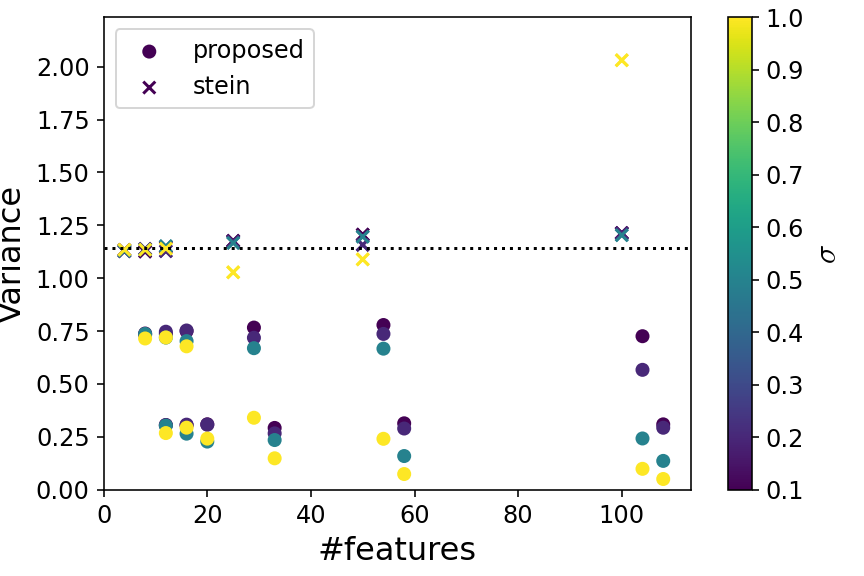}
\caption{The residual variance is plotted against the number of feature functions in the linear model (\eq{eq:linear-model}). The cross marks and circles represent the cases where only the functions based on Stein's identity are used, and where the proposed ratio-based functions are combined with them, respectively. The dotted line shows the variance of the target function $\var{\vx\sim p}{\ftar(\vx)}$.}
\label{fig:fig2c}
\end{figure}

We plot the residual variance against the number of feature functions used for the linear model in Fig.~\ref{fig:fig2c}, varying the parameters as $\sigma \in\{0.1, 0.2, 0.5, 1.0\}$, $\mst = \{4, 8, 12, 25, 50, 100\}$, and $\mra \in \{4, 8\}$.
Regardless of these parameter choices, the combination of the two classes of functions leads to a quantitatively better linear approximation that will achieve a smaller variance for estimating the expectation $I$.

\section{Discussion}

In this paper, we proposed a ratio-based construction of zero-mean feature functions for variance reduction in MC estimation of expectations.
Since our numerical demonstration is limited to a synthetic two-dimensional case, how the performance is retained in more realistic cases in higher-dimensional spaces remains a topic for future work.
In addition, the procedure for choosing the reference mixture $R$ is not limited to the one used in the numerical example, and there is room for improvement.

As mentioned in the earlier section, the zero-mean feature functions can be used to design the herding dynamics \cite{Herding,Herding2010,NOLTA2025} for improving sample generation for the MC method. 
Applying the proposed method to this design and evaluating its performance is also an interesting direction for future exploration.

\section*{Acknowledgments}
This work was supported by JSPS KAKENHI Grant Number 25K21300, JST ALCA-Next Grant No. JPMJAN23F2, JST Moonshot R\&D Grant No. JPMJMS2021, JST CREST Grant No. JPMJCR25R1, and a project, JPNP14004, commissioned by the New Energy and Industrial Technology Development Organization (NEDO).

\end{document}